\documentclass[aps,showpacs,preprintnumbers,amsmath,amssymb,preprint]{revtex4}
\usepackage{graphicx}

\begin{document}

\title{\hfill {\small\bf ECTP-2009-5}\\
Viscous Quark-Gluon Plasma in the Early Universe}

\author{ A. Tawfik$^{1}$, M. Wahba$^1$, H. Mansour$^2$ and T. Harko$^3$ \\ 
{\small $^1$Egyptian Center for Theoretical Physics (ECTP), MTI University,
 Cairo-Egypt} \\
{\small $^2$Department of Physics, Faculty of Science, Cairo University, Giza-Egypt} \\
 {\small $^3$Department of Physics and Center for Theoretical
and Computational Physics,}\\ {\small University of Hong Kong, Hong Kong}\\
}

\date{\today}

\begin{abstract}

We consider the evolution of a flat, isotropic and homogeneous
Friedmann-Robertson-Walker Universe, filled with a causal bulk viscous
cosmological fluid, that can be characterized by an ultra-relativistic
equation of state and bulk viscosity coefficient obtained from recent lattice
QCD calculations. The basic equation for the Hubble parameter is derived under
the assumption that the total energy in the Universe is conserved. By assuming
a power law dependence of bulk viscosity coefficient, temperature and
relaxation time on energy density, an approximate solution of the field
equations has been obtained, in which we utilized equations of state from recent
lattice QCD simulations QCD and heavy-ion collisions to derive an evolution equation. 
In this treatment for the viscous cosmology, we found no evidence for singularity. For example, both Hubble parameter 
and scale factor are finite at $t=0$, $t$ is the comoving time. Furthermore, 
their time evolution essentially differs from the one associated with non-viscous and 
ideal gas. Also thermodynamic quantities, like temperature, energy density and 
bulk pressure remain finite as well. In order to prove that the free parameter in our model 
does influence the final results, qualitatively, we checked out other particular solutions. 

\end{abstract}


\maketitle

\section{Introduction}\label{sec:intro}

The dissipative effects, including both bulk and shear viscosity, are supposed
to play a very important role in the early evolution of the Universe. The
first attempts at creating a theory of relativistic fluids were those of
Eckart \cite{Ec40} and Landau and Lifshitz \cite{LaLi87}. These theories are
now known to be pathological in several respects. Regardless of the choice
of equation of state, all equilibrium states in these theories are unstable
and in addition signals may be propagated through the fluid at velocities
exceeding the speed of light. These problems arise due to the first order
nature of the theory, that is, it considers only first-order deviations from
the equilibrium leading to parabolic differential equations, hence to
infinite speeds of propagation for heat flow and viscosity, in contradiction
with the principle of causality. Conventional theory is thus applicable only
to phenomena which are quasi-stationary, i.e. slowly varying on space and
time scales characterized by mean free path and mean collision time.

A relativistic second-order theory was found by Israel \cite{Is76} and
developed by Israel and Stewart \cite{IsSt76}, Hiscock and Lindblom \cite%
{HiLi89} and Hiscock and Salmonson \cite{HiSa91} into what is called
``transient'' or ``extended'' irreversible thermodynamics. In this model
deviations from equilibrium (bulk stress, heat flow and shear stress) are
treated as independent dynamical variables, leading to a total of 14
dynamical fluid variables to be determined. For general reviews on causal
thermodynamics and its role in relativity see \cite{Ma95}.

Causal bulk viscous thermodynamics has been extensively used for describing
the dynamics and evolution of the early Universe or in an astrophysical
context. But due to the complicated character of the evolution equations,
very few exact cosmological solutions of the gravitational field equations
are known in the framework of the full causal theory. For a homogeneous
Universe filled with a full causal viscous fluid source obeying the relation
$\xi \sim \rho ^{1/2}$, with $\rho $ the energy density of the cosmological
fluid, exact general solutions of the field equations have been obtained in %
\cite{ChJa97,MaHa99a,MaHa99b,MaHa00a,MaTr97}. It has also been proposed that causal bulk
viscous thermodynamics can model on a phenomenological level matter creation
in the early Universe \cite{ChJa97}. Exact causal viscous
cosmologies with $\xi \sim \rho ^{s},s\neq 1/2$ have been considered in Ref.~\cite{MaHa99a}.

Because of technical reasons, most investigations of dissipative causal
cosmologies have assumed Friedmann-Robertson-Walker (FRW) symmetry (i.e. homogeneity and isotropy) or
small perturbations around it \cite{MaTr97}. The Einstein field equations
for homogeneous models with dissipative fluids can be decoupled and
therefore are reduced to an autonomous system of first order ordinary
differential equations, which can be analyzed qualitatively \cite{CoHo95}.

The role of a transient bulk
viscosity in a FRW space-time with decaying vacuum has been discussed in %
\cite{AbVi97}. Models with causal bulk viscous cosmological fluid have been
considered recently~\cite{ArBe00}. They obtained both
power-law and inflationary solutions, with the gravitational constant an
increasing function of time. The dynamics of a viscous cosmological fluids in
the generalized Randall-Sundrum model for an isotropic brane were considered
in \cite{Chen01}. The renormalization group method was applied to the study of
homogeneous and flat FRW Universes, filled with a causal bulk viscous cosmological
fluid, in \cite{Be03}. A generalization of the Chaplygin gas model, by assuming
the presence of a bulk viscous type dissipative term in the effective thermodynamic
pressure of the gas, was investigated recently in \cite{Pun08}.

Recent RHIC results give a strong indication that in the heavy-ion collisions
experiments, a hot dense matter can be formed~\cite{reff1}. Such an experimental
evidence might agree with the "new state of matter"
as predicted in the Lattice QCD
simulations~\cite{reff5}. However, the experimentally observed elliptic flow in
peripheral heavy-ion collisions seems to indicate that a thermalized collective QCD
matter has been produced. In a addition to that, the success of ideal fluid dynamics
in explaining several experimental data e.g. transverse momentum spectra of
identified particles, elliptic flow~\cite{reff6}, together with the string theory
motivated that the shear viscosity $\eta$ to the entropy $s$ would have the lower
limit $\approx 1/4\pi$~\cite{reff7} leading to a paradigm that in heavy-
ion collisions, that a {\it nearly} perfect fluid likely be created and the quarks
and gluons likely go through relatively rapid equilibrium characterized with
a thermalization time less than $1$~fm/c~\cite{mueller1}.

According to recent lattice QCD simulations~\cite{mueller2}, the bulk viscosity $
\xi $ is not negligible near the QCD critical temperature $T_c$. It has been shown
that the bulk and shear viscosity at high temperature $T$ and weak coupling $\alpha_s
$, $\xi\sim  \alpha_s^2 T^3/\ln \alpha_s^{-1}$ and $\eta\sim T^3/(\alpha_s^2 \ln
\alpha_s^{-1})$~\cite{mueller3}. Such a behavior obviously reflects the fact that
near $T_c$ QCD is far from being conformal. But at high $T$, QCD approaches
conformal invariance, which can be indicated by low trace anomaly $(\epsilon-3p)/T^4$
~\cite{karsh09}, where $\epsilon$ and $p$ are energy and pressure density,
respectively. In the quenched lattice QCD, the ratio $\zeta/s$ seems to diverge near
$T_c$~\cite{meyer08}.

To avoid the mathematical difficulties accompanied with the Abel second type non-homogeneous
and non-linear differential equations~\cite{TawCosmos}, one used to model the cosmological fluid as an ideal
(non-viscous) fluid. No doubt that the viscous treatment of the cosmological background
should have many essential consequences~\cite{taw08}. The thermodynamical ones, for instance,
can profoundly modify the dynamics and configurations of the whole cosmological background~\cite{conseq1}.
The reason is obvious. The bulk viscosity is to be expressed as a function of the
Universe energy density $\rho$~\cite{conseq2}. Much progress has been achieved in relativistic
thermodynamics of dissipative fluids. The pioneering theories of Eckart~\cite{Ec40} and
Landau and Lifshitz~\cite{LaLi87} suffer from lake of causality constrains. The currently
used theory is the Israel and Stewart theory~\cite{Is76,IsSt76}, in which the causality is
conserved and theory itself seems to be stable~\cite{HiLi89,Ma95}.

In this article, we aim to investigate the effects that bulk viscosity has on the
Early Universe. We consider a background corresponding to a
FRW model filled with ultra-relativistic viscous
matter, whose bulk viscosity and equation of state have been deduced from
recent heavy-ion collisions experiments and lattice QCD simulations.

The present paper is organized as follows. The basic equations of the model
are written down in Section~\ref{field}. In Section~\ref{approx} we present an approximate solution of the
evolution equation. Section~\ref{part1} is devoted to one particular solution, in which we assume that $H=const.$ The results and conclusions are given in Sections~\ref{final} and ~\ref{final2}, respectively.

\section{Evolution equations}\label{field}

We assume that geometry of the early Universe is filled with a bulk viscous
cosmological fluid, which can be described by a spatially flat FRW type metric
given by
\begin{equation}  \label{1}
ds^{2}=dt^{2}-a^{2}\left( t\right) \left[ dr^{2}+r^{2}\left( d\theta
^{2}+\sin ^{2}\theta d\phi ^{2}\right) \right] .
\end{equation}
The Einstein gravitational field equations are:
\begin{equation}
R_{ik}-\frac{1}{2}g_{ik}R=8\pi GT_{ik}.  \label{ein}
\end{equation}
In rest of this article, we take into consideration natural units, i.e., $c=1$, for instance.

The energy-momentum tensor of the bulk viscous cosmological fluid filling the
very early Universe is given by
\begin{equation}
T_{i}^{k}=\left(  \rho +p+\Pi\right)  u_{i}u^{k}-\left(  p+\Pi\right)
\delta_{i}^{k},\label{1_a}%
\end{equation}
where $i,k$ takes $0,1,2,3$, $\rho$ is the mass density, $p$ the thermodynamic pressure, $\Pi $ the
bulk viscous pressure and $u_{i}$ the four velocity satisfying the condition
$u_{i}u^{i}=1$. The particle and entropy fluxes are defined according to
$N^{i}=nu^{i}$ and $S^{i}=sN^{i}-\left(  \tau\Pi^{2}/2\xi T\right)
u^{i}$, where $n$ is the number density, $s$ the specific entropy, $T\geq0$ the
temperature, $\xi$ the bulk viscosity coefficient, and $\tau\geq0$ the
relaxation coefficient for transient bulk viscous effect (i.e. the relaxation time), respectively.

The evolution of the cosmological fluid is subject to the dynamical laws of
particle number conservation $N_{\text{ };i}^{i}=0$ and Gibbs' equation
$Td\rho=d\left(  \rho /n\right)  +pd\left(  1/n\right)  $.
In the following we shall also suppose that the energy-momentum tensor of the
cosmological fluid is conserved, that is $T_{i;k}^{k}=0$.

The bulk viscous effects can be generally described by means of an effective
pressure $\Pi $, formally included in the effective thermodynamic pressure $%
p_{eff}=p+\Pi $ \cite{Ma95}. Then in the comoving frame the energy momentum tensor
has the components $T_{0}^{0}=\rho ,T_{1}^{1}=T_{2}^{2}=T_{3}^{3}=-p_{eff}$.
For the line element given by Eq.~(\ref{1}), the Einstein field equations read
\begin{eqnarray}  \label{2}
\left( \frac{\dot{a}}{a}\right)^{2} &=& \frac{8\pi}{3}G \;\rho, \\
\frac{\ddot{a}}{a} &=& -\frac{4\pi}{3}G \; \left( 3p_{eff}+\rho \right),
\label{3}
\end{eqnarray}
where one dot denotes derivative with respect to the time $t$, $G$ is the gravitational constant and $a$ is the scale factor.

Assuming that the total matter content of the Universe is conserved, $%
T_{i;j}^j=0$, the energy density of the cosmic matter fulfills the
conservation law:
\begin{equation}  \label{5}
\dot{\rho}+3H\left( p_{eff}+\rho \right) =0,
\end{equation}
where we introduced the Hubble parameter $H=\dot{a}/a$.
In presence of bulk viscous stress $\Pi $, the effective
thermodynamic pressure term becomes $p_{eff}=p+\Pi $. Then Eq.~(\ref{5}) can be written as
\begin{equation}  \label{6}
\dot{\rho}+3H\left( p+\rho \right) =-3\Pi H.
\end{equation}

For the evolution of the bulk viscous pressure we adopt the causal evolution
equation \cite{Ma95}, obtained in the simplest way (linear in $\Pi)$ to
satisfy the $H$-theorem (i.e., for the entropy production to be non-negative,
$S_{;i}^{i}=\Pi^{2}/\xi T\geq0$ \cite{Is76,IsSt76}). According to the causal relativistic
Israel-Stewart theory, the evolution equation of the bulk viscous pressure reads~\cite{Ma95}
\begin{equation}  \label{8}
\tau \dot{\Pi}+\Pi =-3\xi H-\frac{1}{2}\tau \Pi \left( 3H+\frac{\dot{\tau}}{%
\tau }-\frac{\dot{\xi}}{\xi }-\frac{\dot{T}}{T}\right).
\end{equation}
In order to have a closed system from equations (\ref{2}), (\ref{6}) and (\ref{8})
we have to add the equations of state for $p$ and $T$.

As shown in Appendix~A, the equation of state, the temperature and the bulk
viscosity of the quark-gluon plasma (QGP), can be determined approximately at high temperatures~\cite{karsch07} from recent lattice QCD calculations~\cite{Cheng:2007jq}, as
\begin{equation}\label{13}
P = \omega \rho,\hspace*{1cm}T = \beta \rho^r,\hspace*{1cm}\xi = \alpha \rho + \frac{9}{\omega_0} T_c^4,
\end{equation}
with $\omega = (\gamma-1)$, $\gamma \simeq 1.183$, $r\simeq 0.213$, $\beta\simeq 0.718$,
\begin{equation}
\alpha = \frac{1}{9\omega_0}  \frac{9\gamma^2-24\gamma+16}{\gamma-1},
\end{equation}
and  $\omega_0 \simeq 0.5-1.5$ GeV. In the following we assume that $\alpha \rho >> 9/\omega_0 T_c^4$, and therefore we take $\xi \simeq \alpha \rho$. In order to close the system of the cosmological equations, we have also to give the expression of the relaxation time $\tau $, for which we adopt the expression \cite{Ma95},
\begin{equation}\label{tau}
\tau=\xi\rho^{-1}\simeq\alpha .
\end{equation}

Eqs.~(\ref{13})
are standard in the study of the viscous cosmological models, whereas the equation for $\tau$ is a
simple procedure to ensure that the speed of viscous pulses does not exceed
the speed of light. Eq. (\ref{tau}) implies that the relaxation time in our treatment is constant but strongly depends on EoS. These equations are without sufficient thermodynamical motivation,
but in the absence of better alternatives, we shall follow the practice of adopting
them in the hope that they will at least provide some indication of the range of
bulk viscous effects. The temperature law is the simplest law guaranteeing positive
heat capacity.

With the use of Eqs.~(\ref{8}), (\ref{13}) and (\ref{tau}), respectively, we obtain the following equation describing the cosmological evolution of the Hubble function $H$
\begin{eqnarray}\label{init}
\ddot H + \frac{3}{2} [1+(1-r) \gamma] H\dot H + \frac{1}{\alpha}\dot H - %\nonumber\\
(1+r) H^{-1} \dot H^2 + \frac{9}{4}(\gamma -2) H^3 +
\frac{3}{2}\frac{\gamma}{\alpha} H^2 &=& 0.
\end{eqnarray}

\section{An approximate solution}\label{approx}

We introduce the transformation $u=\dot{H}$, so that Eq.~(\ref{init}) is transformed into a first order ordinary differential equation,
\begin{equation}\label{init2}
u\frac{du}{dH}-(1+r)H^{-1}u^{2}+\left(\frac{3}{2}[1+(1-r)\gamma ]H+\alpha
^{-1}\right) u+\frac{9}{4}\frac{1}{(\gamma)}H^{3}+\frac{3}{2}\frac{\gamma}{\alpha}
H^{2}=0.
\end{equation}
We can rewrite Eq.~(\ref{init2}) in the form
\begin{equation} \label{OmegH1}
\Omega \frac{d\Omega }{dH} = F_1(H)\Omega + F_0(H),
\end{equation}
where
\begin{eqnarray}
\Omega &=& u \; E \;\; = u\; \exp\left(-\int \frac{1+r}{H} dH\right), \nonumber \\
F_1(H) &=& -\left( \frac{3}{2} [1+(1-r)\gamma]H + \frac{1}{\alpha}\right)E, \nonumber \\
F_0(H) &=& -\left(\frac{9}{4}(\gamma-2) H^3 + \frac{3}{2} \frac{\gamma}{\alpha} H^2\right)E^2. \nonumber
\end{eqnarray}
By introducing a new independent variable $z=\int F_1(H)\,dH$, we obtain
\begin{equation}
\Omega \frac{d\Omega}{dz} - \Omega = g(z),
\end{equation}
with $g(z)$ is defined parametrically as,
\begin{equation} \label{gofzz1}
g(z) = \frac{F_0}{F_1}.
\end{equation}
As shown in Appendix~B, %\ref{App:A}, 
$g(z)$ can be approximated as a simple function of $z$
\begin{equation}
g(z)\approx {\cal C}\; z,
\end{equation}
where ${\cal C}$ is a constant. We proceed with this approximation to get
solvable differential equations. Keeping the parametric solution of $g(z)$,
Eq.~(\ref{fullgofz}), results in much more complicated differential
equations. This would be the subject of a future work.

From the definitions of $\Omega$ and $z$  we have
\begin{eqnarray}
\Omega &=& H^{1+r}\dot H, \label{Eq1} \\
z &=& H^{2+r} \left(\frac{-3[1+(1-r)\gamma]H}{2(1-r)} +\frac
{1}{\alpha r}\right), \label{Eq2}
\end{eqnarray}
Analogous to the solution of reduced Abel type canonical equation,
\begin{equation}\label{abel1}
y \frac{dy}{dx} - y = a x
\end{equation}
(see Appendix C) %.~\ref{App:B}), 
we obtain the relation $\Omega = z/{\cal P}$. Therefore, from Eqs.~(\ref{Eq1})
and (\ref{Eq2}) we obtain the following first order differential equation for Hubble parameter $H$,
\begin{equation} \label{init-polyn}
{\cal P} \dot H = \frac{-3[1+(1-r)\gamma]}{2(1-r)} H^2 +\frac{1}{\alpha r}H
\end{equation}
with the solution
\begin{equation}
H(t) = \frac{B}{\exp(-Bt/{\cal P})-A} \label{eq:mysolut1}
\end{equation}
where
\begin{equation}\label{paramsab}
A=\frac{-3[1+(1-r)\gamma]}{2(1-r)}, \hspace*{1cm} B=\frac{1}{\alpha r},
\end{equation}
and ${\cal P}$ is taken as a free parameter. We can assign any real value to ${\cal P}$. For the results presented in this work, we used a negative value. This negative sign is necessarily to overcome the sign from the integral limits. The geometric and thermodynamic
quantities of the Universe read
\begin{eqnarray}
 a(t) &=& a_0\left(\frac{\exp(-B t/{\cal P})}{\exp(-B
   t/{\cal P})-A}\right)^{{\cal P}/A} \label{approx-a}, \\
\rho(t) &=& 3\, H^2= 3 \left(\frac{B}{\exp(-Bt/{\cal P})-A}\right)^2 \label{approx-rho},\\
T(t) &=& \beta \rho^{r}=\beta \left(3 \frac{B^{2}}{[\exp(-Bt/{\cal P})-A]^{2}}\right)^r \label{approx-T},\\
\Pi(t) &=& -2\dot{H}-3\gamma H^{2}=-\frac{B^2}{{\cal P}} \left(\frac{2\exp(-Bt/{\cal P})+3\gamma{\cal P}}{[\exp(-Bt/{\cal P})-A]^2}\right) \label{approx-Pi},\\
 q(t)&=&\frac{d}{dt}H^{-1}-1=-\frac{1}{{\cal P}}\exp(-Bt/{\cal P})-1. \label{approx-q}
\end{eqnarray}
$a_0$ is an arbitrary constant of the integration.
The sign of $q$ indicates whether the Universe decelerates (positive) or
accelerates (negative). $q$ can also be given as a function of the
thermodynamic, gravitational and cosmological quantities $q(t)=[\rho(t)
  +3p(t)+3\Pi(t)]/2\rho(t)$~\cite{kolbBook}.

\section{de Sitter Universe }\label{part1}

Besides the approximation in $g(z)$, previous solution apparently depends on
the free parameter ${\cal P}$. In this section, we suggest a particular solution to
overcome ${\cal P}$. Eq.~(\ref{init}) can easily be obtained by assuming that $H$ doesn't depend one $t$, i.e, de Sitter Universe. With a simple calculation, we get an estimation for $H$
\begin{equation}\label{partcH}
H=\frac{4}{9}\frac{\alpha ^{-1}\gamma}{2-\gamma }.
\end{equation}
The geometric and thermodynamic parameters of the Universe are given by
\begin{eqnarray}
a(t) &=& a_{0}\exp \left[ \frac{4\alpha ^{-1}\gamma }{9(2-\gamma )}t\right], \label{partca} \\
\rho(t) &=& 3\left[ \frac{4\alpha ^{-1}\gamma }{9(2-\gamma )}\right] ^{2}, \label{partcrho}\\
T(t) &=& 3^{r}\beta \left[ \frac{4\alpha ^{-1}\gamma }{9(2-\gamma )}\right]
^{2r}, \label{partcT}\\
\Pi(t) &=& -3\gamma \left[ \frac{4\alpha
^{-1}\gamma }{9(2-\gamma )}\right] ^{2}, \label{partcPi}\\
q(t) &=& -1. \label{partcq}
\end{eqnarray}
Although we have assumed here that the cosmic background is filled with viscous
matter, the assumption that $H=const$ results in an exponential scale
parameter, Eq.~(\ref{partca}). This behavior characterizes the de Sitter
space, when $\Lambda=k=0$. $\rho$ and $T$ are finite at small $t$ as given in Fig.~\ref{Figg2}.

\section{Particular Solution}\label{part2}

Another particular solution for Eq.~(\ref{init}) can be obtained, when
assuming that the dependence of $u$ on $H$ can be given by the polynomial
in Eq.~(\ref{init-polyn})
\begin{equation}\label{init-partc2}
u=b_{1}H^{2}+b_{2}H,
\end{equation}
where $b_{1}$ and $b_{2}$ are constants. Some simple calculations show that
this form is a solution of the initial equation, Eq.~(\ref{init2}), if
\begin{eqnarray}
b_1 &=&-\frac{3}{2}\frac{1+\gamma }{1-r}, \\
b_2 &=& \frac{1}{r\alpha }.
\end{eqnarray}
$b_2$ is identical to $B$ in Eq.~(\ref{paramsab}). $r$ and $\gamma$ have to satisfy the compatibility relation
\begin{equation}
r=\frac{2-\gamma}{2+\gamma^2}.
\end{equation}
Integrating Eq.~(\ref{init-partc2}) results in
\begin{eqnarray} \label{Eq:Ht}
H(t) &=& \frac{b_2\exp(-b_2 t)}{1-b_1\exp(-b_2 t)},
\end{eqnarray}
where minus sign in the exponential function refers to flipping the integral limits. This was not necessary while deriving the expressions given in Section~\ref{approx}. The free parameter {\cal P} compensates it.
The geometric and thermodynamic quantities of the Universe read
\begin{eqnarray}
a(t)&=&a_0\left(\frac{\exp(b_2t)-b_1}{\exp(b_2t)}\right)^{1/b_1}, \label{partc2a} \\
\rho(t)&=& 3 \left(\frac{b_2\exp(-b_2 t)}{1-b_1\exp(-b_2 t)}\right)^2, \label{partc2rho}\\
T(t)&=& 3^r\;\beta \left(\frac{b_2\exp(-b_2 t)}{1-b_1\exp(-b_2 t)}\right)^{2r}, \label{partc2T}\\
\Pi(t)&=& \frac{b_2^2 \left[2\exp(b_2t)-3\gamma\right]}{\left[\exp(b_2t)-b_1\right]^2}, \label{partc2Pi}\\
q(t) &=& \exp(b_2t)-1. \label{partc2q}
\end{eqnarray}
Obviously , we notice that the scale parameter in Eq.~(\ref{partc2a}) looks
like Eq.~(\ref{approx-a}), which strongly depends on the free parameter ${\cal
  P}$. The other geometric and thermodynamic quantities find similarities in
Eq.~(\ref{approx-rho})~-~(\ref{approx-q}), respectively. Deceleration
parameter $q$ seems to be positive everywhere.

\section{Results}\label{final}

In present work, we have considered the evolution of a full causal bulk
viscous flat, isotropic and homogeneous Universe with bulk viscosity parameters and equation of state taken from recent lattice QCD data and heavy-ion collisions.
Three classes of solutions of the evolution equation have been obtained.

\begin{figure}
\includegraphics[width=5cm,angle=-90]{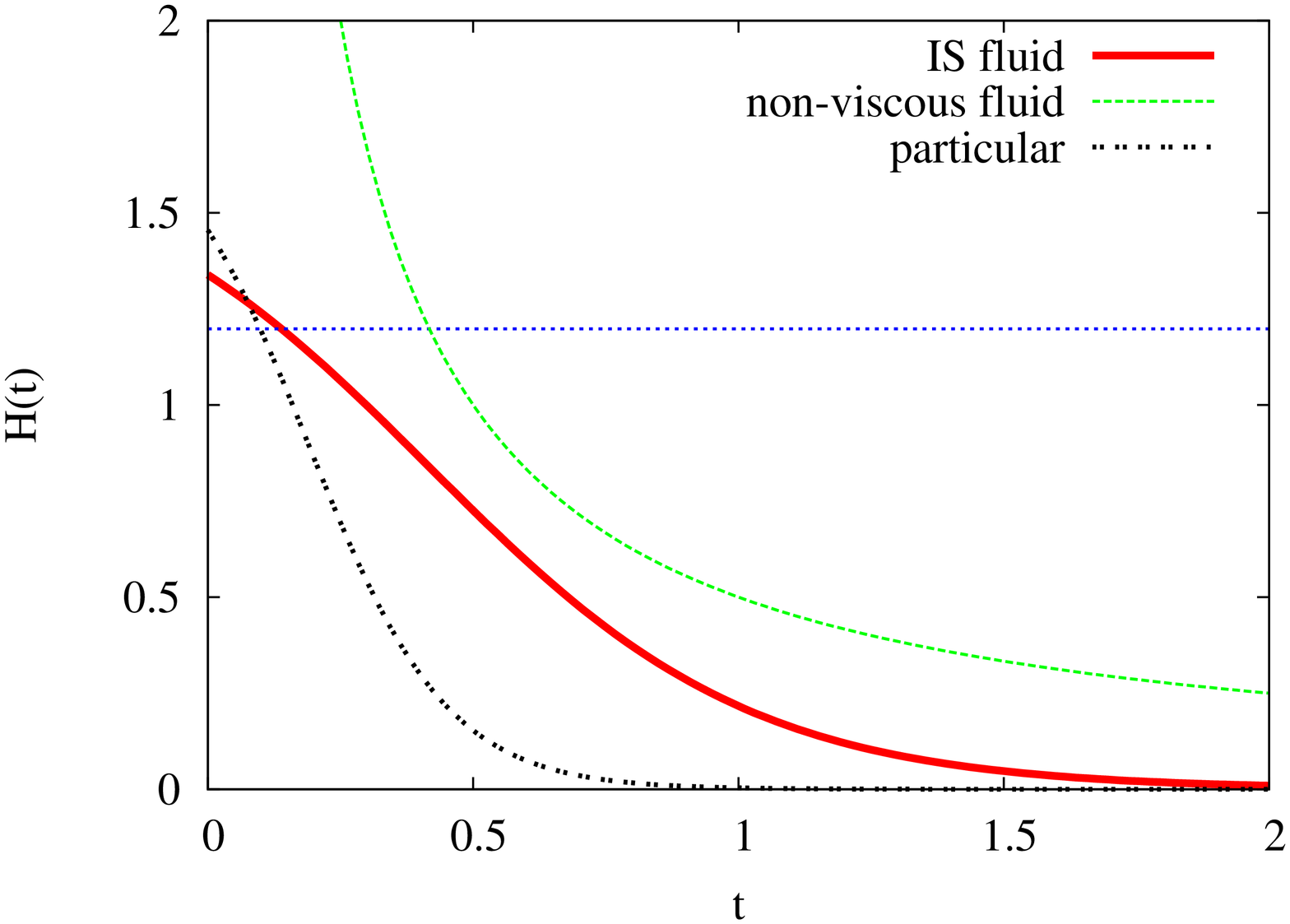}
\includegraphics[width=5cm,angle=-90]{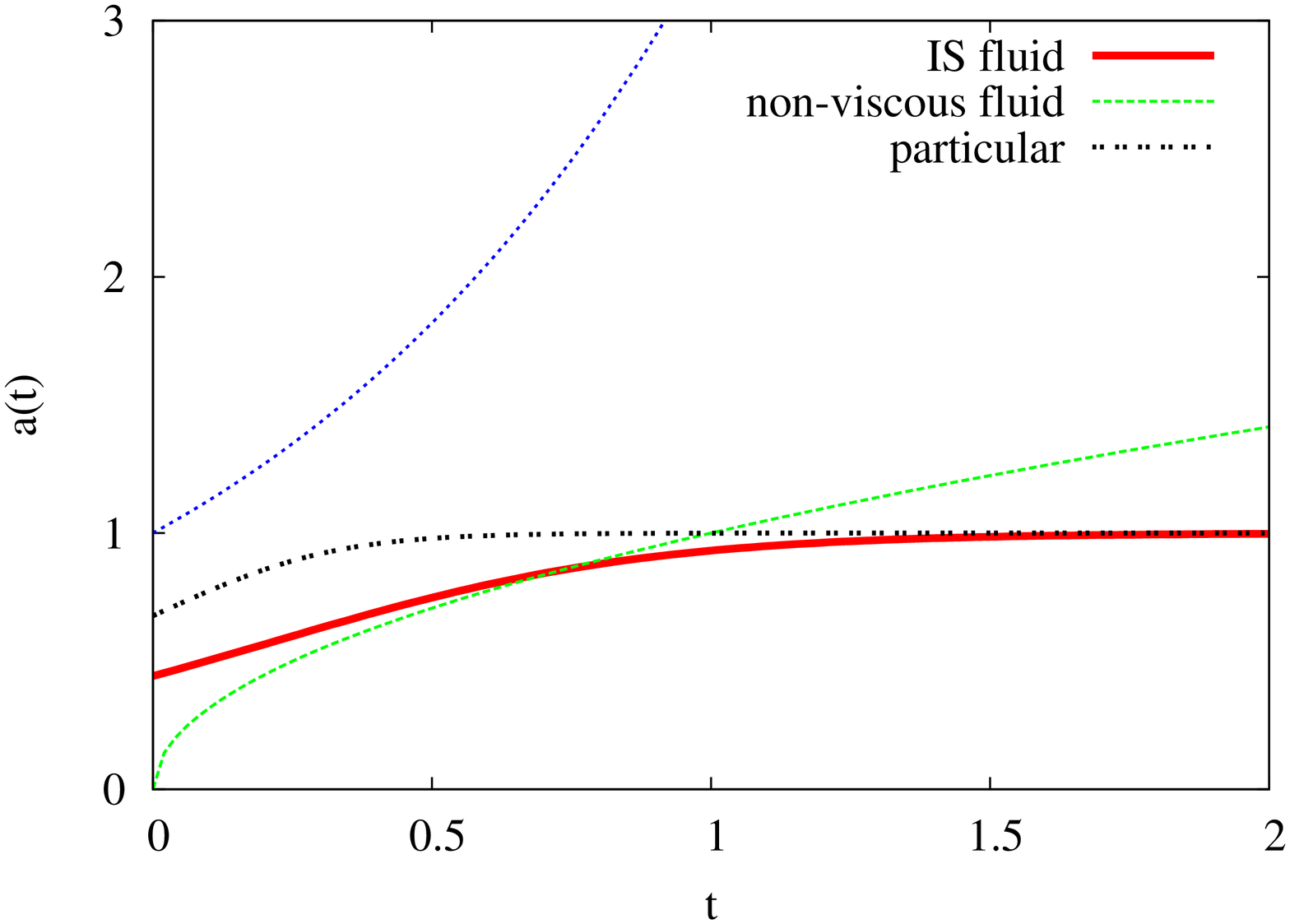}
\caption{Left panel: $H(t)$ vs. $t$ for viscous (solid line) and non-viscous
  fluid (dashed line). Contrary to non-viscous fluid, $H(t)$ is our solution
  shows no singularity. The straight line gives the results from
  the particular solution, Eq.~(\ref{partcH}). Double dotted curve represent
  the particular solution, Eq.~(\ref{Eq:Ht}). Right panel: $a(t)$
  vs. $t$. The approximate solution gives finite $a(t)$ at $t=0$. (solid
  line). Dotted straight line represents the results from
  Eq.~(\ref{partca}). The particular solution is given by the double dotted
  line, Eq.~(\ref{partc2a}). }
\label{Figg1}
\end{figure}

In Fig.~\ref{Figg1}, $H(t)$ and $a(t)$ are depicted in dependence on the
  comoving time $t$. We compare $H(t)$, given by Eq.~(\ref{eq:mysolut1}), and
  $a(t)$, given by Eq.~(\ref{approx-a}),
with the counterpart parameters obtained in the case when the background matter is
assumed to be an ideal and non-viscous fluid, described by the equations of
  state of the non-interacting ideal gas,
\begin{eqnarray}
H(t) &=& \frac{1}{2t} \label{htideal1}, \\
a(t) &=& \sqrt{t}. \label{atideal1}
\end{eqnarray}

\begin{figure}
\includegraphics[width=5cm,angle=-90]{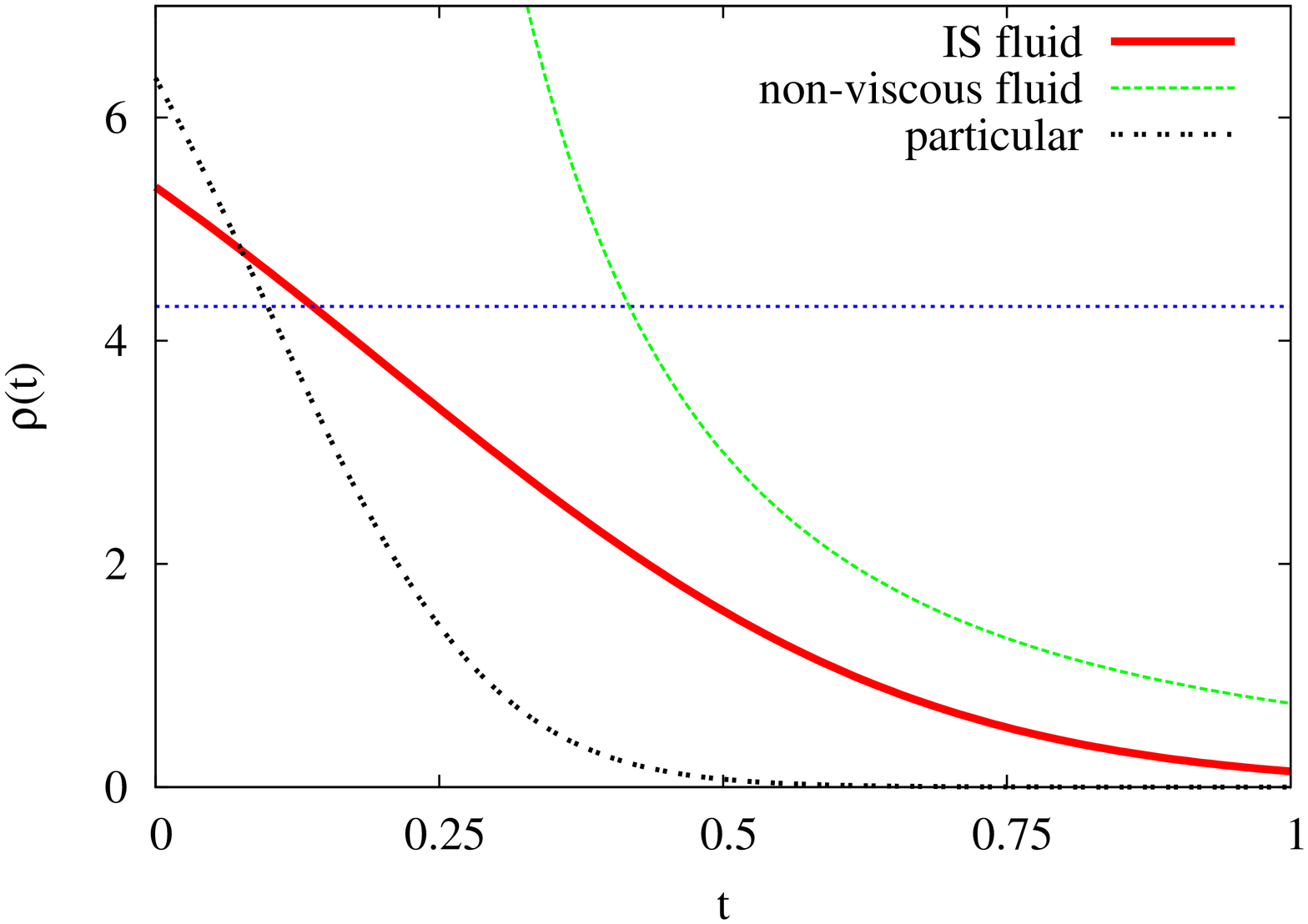}
\includegraphics[width=5cm,angle=-90]{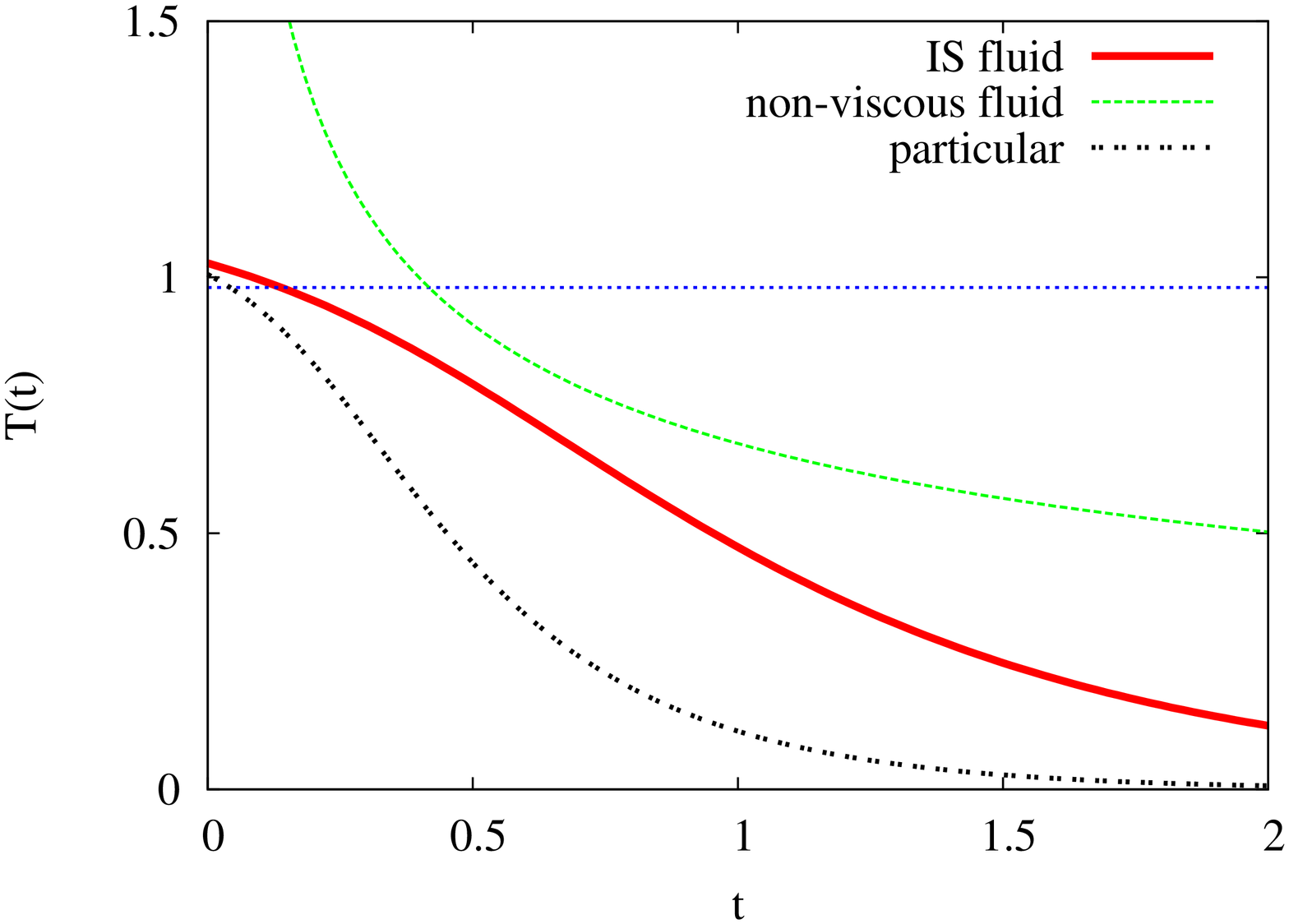}
\caption{Energy density $\rho(t)$ as a function of $t$ (left
panel). The dependence of $T$ on $t$ is given in the right
panel. The curves are as in Fig.~\ref{Figg1}. In both case, viscous fluid
gives no singularity at vanishing $t$. Straight curve are from
Eq.~(\ref{partcrho}) and (\ref{partcT}), respectively.}
\label{Figg2}
\end{figure}

\begin{figure}
\includegraphics[width=5cm,angle=-90]{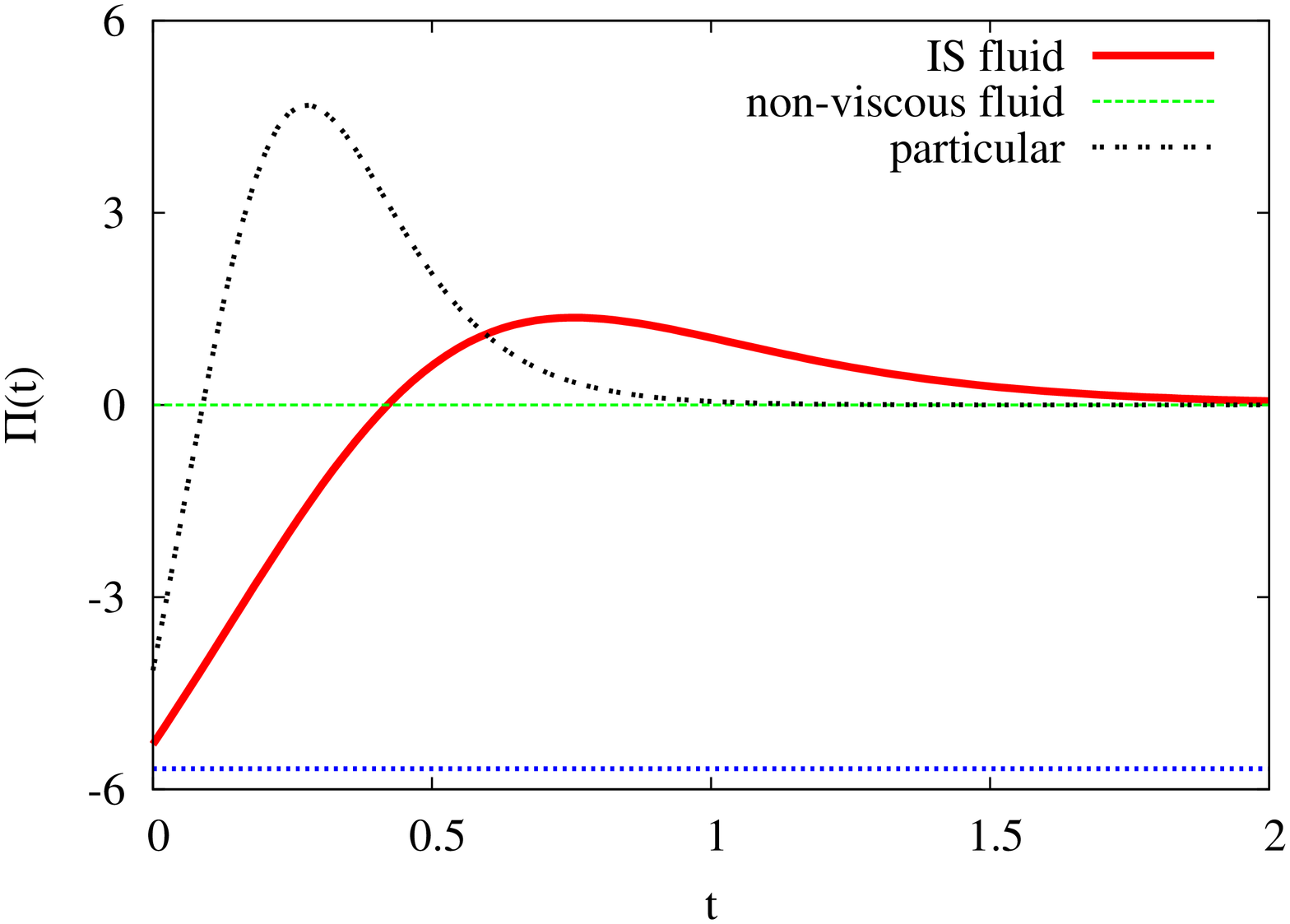}
\includegraphics[width=5cm,angle=-90]{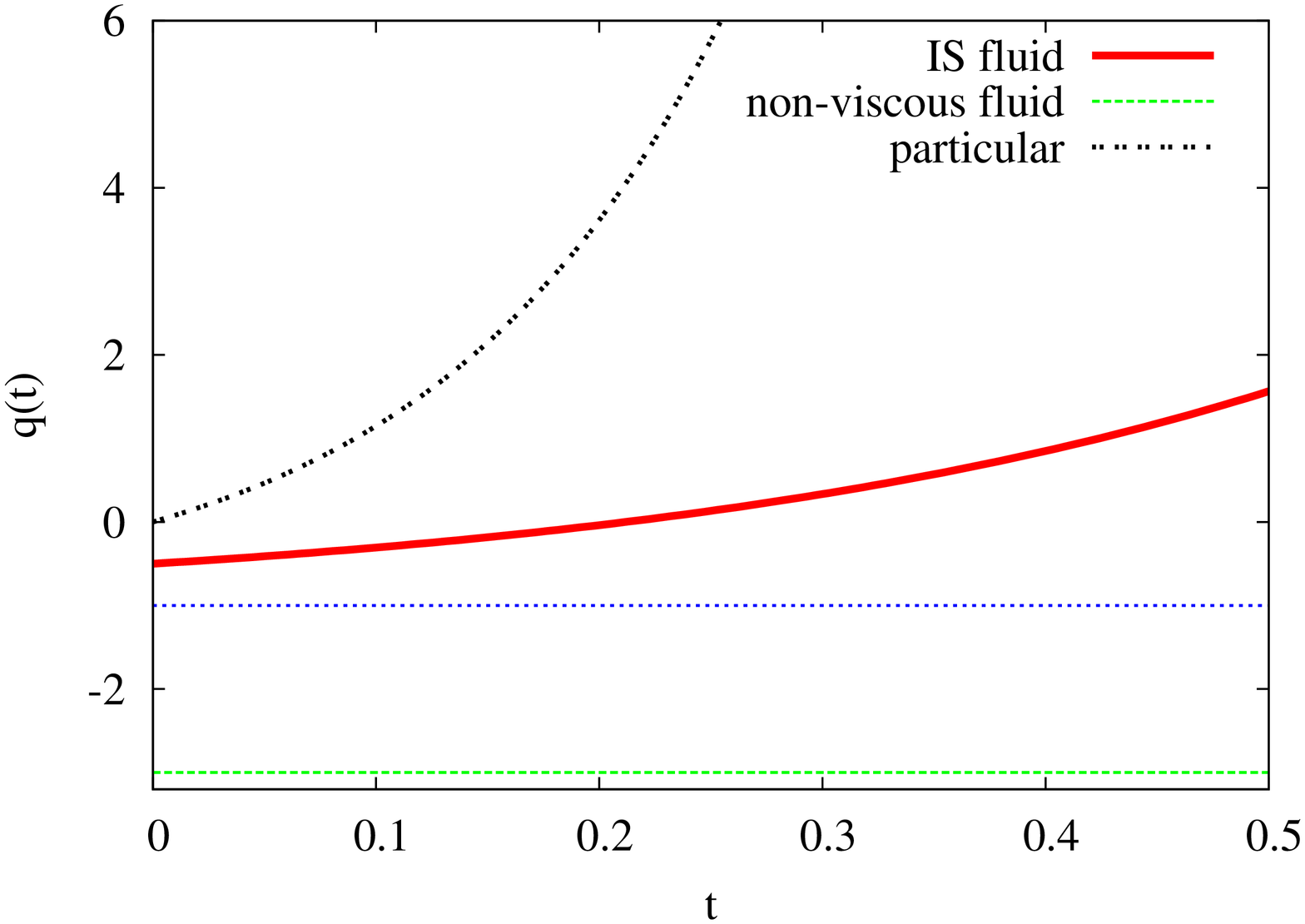}
\caption{Left panel: bulk pressure $\Pi$ depicted in dependence on
  $t$. At small $t$, $\Pi$ jumps from negative to positive values. At larger
  $t$, $\Pi$ vanishes as the case in the non-viscous fluid (dashed line). The
  bottom line gives $\Pi$ according to first particular solution,
  Eq.~(\ref{partcPi}). Right panel: deceleration parameter $q$ is depicted
  with $t$. Straight and dotted lines represent non-viscous fluid and first particular
  solution, Eq.~(\ref{partcq}), respectively. Top curve represents second particular solution,
Eq.~(\ref{partc2q}). }
\label{Figg3}
\end{figure}

In the left panel of Fig.~\ref{Figg1}, $H(t)=\dot a/a$ has an
exponential decay, whereas in the non-viscous case, $H(t)$ is decreasing
according to Eq.~(\ref{htideal1}). The latter is much slower than the former,
reflecting the nature of the exponential and linear dependencies. The other
difference between the two cases is obvious at small $t$. We notice a
divergence, or singularity, associated with the ideal non-viscous fluid,
Eq.~(\ref{htideal1}). The viscous fluid results in finite $H$ even at vanishing
$t$, as can be seen from Eq.~(\ref{eq:mysolut1}).

The scale factor $a(t)$ also shows differences in both cases. $a(t)$ in a
Universe with an ideal and non-viscous background matter depends on $t$
according to Eq.~(\ref{atideal1}), which simply implies that $a(t)$ is directly
proportional to $t$, and $a(t)$ vanishes at $t=0$, which shows the existence of a
singularity of $H$. Assuming that the background matter is described
by a viscous fluid results in different $a(t)$-behaviors with increasing
$t$. At $t=0$, $a(t)$ remains finite. Correspondingly, $H(t)$ remains also
finite. In general, the dependence on $t$ is much more complicated than in
Eq.~(\ref{atideal1}). Here we have an $A/{\cal P}$ root of an exponential
function. If $\exp(-Bt/{\cal P})>>A$, $a$ remains constant.

Fig.~\ref{Figg2} illustrates the dependence of the two thermodynamical quantities,
$\rho $ and $T$, on the comoving time. The non-viscous Universe shows
a singular behavior in $\rho$ at vanishing $t$, as shown in the left panel of Fig.~\ref{Figg2}. This is
not obvious in the case where we have taken into consideration a finite viscosity
coefficient, i.e., $\rho$ is finite at $t=0$. In both cases, $\rho$ is
decreasing with increasing $t$, reflecting that the Early Universe was likely
expanding. Also the life time of the thermal viscous Universe seems to be
shorter than for the non-viscous Universe. Almost the same behavior is observed in
the right panel of Fig.~\ref{Figg2}. The temperature $T$ seems to be finite at
vanishing $t$ in the viscous Universe. The $T$-singularity is only present, 
if we assume that the background matter is non-viscous ideal gas.

In left panel of Fig.~\ref{Figg3}, we show the dependence of the bulk viscous pressure $\Pi$ on $t$. $\Pi$ takes
negative values at very small $t$. Then it switches to positive values at some values of $t$. 
After reaching the maximum value, $\Pi$ decays exponentially with
increasing $t$. At larger $t$, $\Pi$ entirely vanishes. The deceleration
parameter $q$, given by Eq.~(\ref{approx-q}), is depicted in the right panel of Fig.~\ref{Figg3}, and it is
compared with $q$ for a  non-viscous fluid, $q=-3$. The approximate solution, given by
Eq.~(\ref{approx-q}), results in negative $q$ at small $t$, referring to
expansion era. $q$ from the particular solution, Eq.~(\ref{partcq}) is negative everywhere.\\

For the particular solution, only the scale factor depends on $t$,
Eq.~(\ref{partca}). The results are given in the right panel of
Fig.~\ref{Figg1}. All cosmological and thermodynamical quantities given by Eq.~(\ref{partcH}) and
Eqs.~(\ref{partcrho})-(\ref{partcq}) are constant in time.

\section{Conclusions}\label{final2}

It is obvious that the bulk viscosity  plays an important role in the
evolution of the Early Universe. Despite of the simplicity of our model, it
shows that a better understanding of the dynamics of our Universe is only
accessible, if we use reliable equation of state to characterize the matter filling out the cosmic background.

We conclude that the causal bulk viscous Universe described by the approximate
solution starts its evolution from an initial non-singular state with a
non-zero initial value of Hubble parameter $H(t)$ and scale factor $a(t)$, where $t$ is the comoving time. In this treatment, $t$ is given in GeV$^{-1}$. Also the
thermodynamical quantities, energy density $\rho$ for instance, are finite at
vanishing $t$. Even the temperature $T$ itself shows no singularity at $t=0$. The Hubble parameter $H$ decreases
monotonically with $T$ similar to $\rho$. The bulk viscous pressure $\Pi$
likely satisfies the condition that $\Pi<0$ at very small $t$ indicating to
inflationary era. Then $\Pi $ switches to positive value. It reaches a maximum
value and then decays and vanishes, exponentially, at large $t$.
The deceleration parameter $q$ shows an expanding behavior in the case of non-viscous ideal gas and first particular solution. For second particular solution, $q$ starts from zero and increases, exponentially. According to this solution, the Universe was decelerating. The approximate solution shows an interesting behavior in $q(t)$, Eq.~(\ref{approx-q}). At small $t$, the values of $q$ are negative, i.e. the Universe was accelerating (expansion). At larger $t$, a non-inflationary behavior sets on, $q>0$, i.e., the Universe switched to a decelerating evolution. 

In this treatment, we assumed that the Universe is flat, $k=0$, and the background geometry is filled out with QCD matter (QGP) with a finite viscosity coefficient. The resulting Universe is obviously characterized by a shortly increasing and afterward constant scale factor and a fast vanishing Hubble parameter. At $t=0$, both $a(t)$ and $H(t)$ remain finite, i.e., there is no singularity. The validity of our treatment depends on the validity of the equations of states, Eq.~\ref{13}, which we have deduced from the lattice QCD simulations at temperatures larger than $T_c\approx 0.19~$GeV. Below $T_c$, as the Universe cooled down, not only the degrees of freedom suddenly increase~\cite{Tawfik03} but also the equations of state turn to be the ones characterizing the hadronic matter. Such a phase transition - from QGP to hadronic matter - would characterize one end of the validity of our treatment. The other limitation is the very high temperatures (energies), at which the strong coupling $\alpha_s$ entirely vanishes.

%%%%%%%%%%%%%%%%%%%%%%%%%%%%%%%%%%%%%

\newpage

\section*{Appendix A: Viscosity coefficient $\xi (T)$ from LQCD}\label{App:C}

Following the discussion presented in {\cite{Cheng:2007jq}}, the bulk viscosity of QGP can be calculated from the lattice QCD by Eq.~(13) in that paper.
We assume that the decay factors for pions and kaons are vanishing above the critical temperature of the phase
transition QGP-hadrons. The quark-antiquark condensates can be neglected at temperatures higher than the critical
one~\cite{TawDom}. Therefore, Eq.~(22) of Ref.~{\cite{Cheng:2007jq}} would be reduced to
\begin{equation} \label{ze}
9\,\omega_0\,\xi = T\, s\, \left(\frac{1}{c_s^2}-3\right)-4(\rho-3p) +16|\epsilon_v|
\end{equation}
where $\rho$ is the energy density and $c_s^2=dp/d\rho $ is the square of the speed of sound.
The parameter $\omega_o$ is a scale depending on the temperature $T$, and defines the validity of the underlying perturbation theory. In this relation, the viscosity is assumed to have a thermal part which can be determined through lattice calculations, and a vacuum contributing part, which can be fixed using quark and gluon condensates. The vacuum part would take the value
\begin{equation}
16 |\epsilon_v| (1 + \frac{3}{8} \cdot 1.6) \simeq (560\ {\rm MeV})^4
\simeq (3 \,T_c)^4 \,
\end{equation}
Our algorithm is the following. Using lattice QCD results on trace anomaly, $(\epsilon-3p)/T^4$, and other thermodynamical quantities, we can determine the bulk viscosity. To make use of the lattice QCD results, it is useful to make a suitable fit to the data at high temperatures.  Then we obtain the following equations of state
\begin{eqnarray}\label{EoS}
p &=&\omega \rho, \hspace*{2cm} %nonumber \\
T =\beta \rho^r, \hspace*{2cm} % \\
c_s^2 = \omega \nonumber
\end{eqnarray}
where $\omega=0.319$, $\beta=0.718\pm 0.054$ and $r=0.23\pm 0.196$.
Using the equations of state, Eq.~(\ref{EoS}) in Eq.~(\ref{ze}), we obtain
\begin{equation}\label{zeta}
 \xi(\epsilon)=\frac{1}{9\omega_o}\frac{9\gamma^2-24\gamma+16}{\gamma-1}\rho+\frac{9}{\omega_o}T_c^4.
\end{equation}

\section*{Appendix B: Estimations of $g(z)$} \label{App:A}

\begin{figure}
\includegraphics[width=8cm,angle=-90]{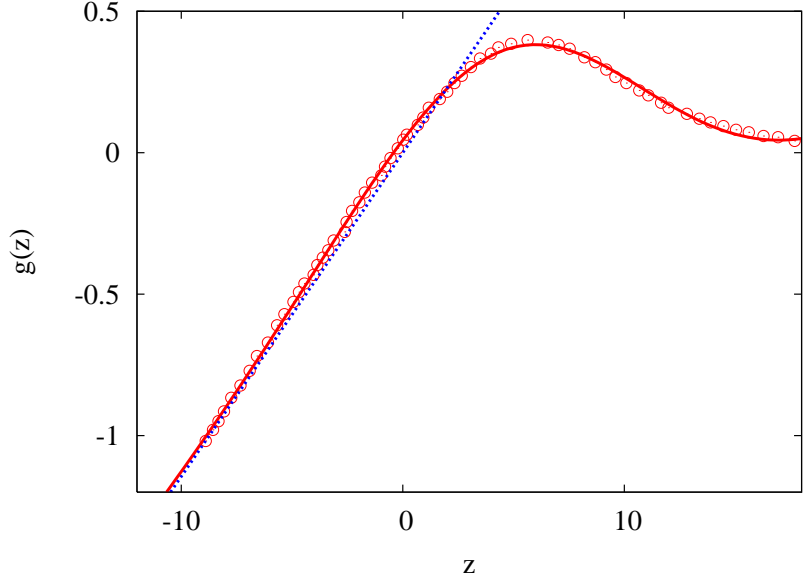}
\caption{The parametric dependence of $g(z)$ on $z$ is given by open symbols. Eq.~(\ref{fullgofz}) is depicted as solid line. The dotted line represents the linear fit,  Eq.~(\ref{lineargofz2}). } 
\label{Figg4}
\end{figure}

For analytical purposes, the function $g(z)$, which is defined in $z$ parameter as $g(z)=F_0/F_1$ in Eq.~(\ref{gofzz1}), can be numerically estimated depending on the parameter $z$  by using the following procedure.
First, we plot it parametrically depending on the parameter $H$, Fig. (\ref{Figg4}). Then we fit the resulting curve to various functions. Based on least-square fit, best choice would be a mixture of polynomial and exponential functions,
\begin{equation} \label{fullgofz}
g(z)= a + b\, z + c \frac{\exp(d\, z)+e}{\left[\exp(d\, z)+f\right]^2}, 
\end{equation}
where the coefficients read $a=-2.078\pm0.117$, $b=0.091\pm0.007$ and $c=17.332\pm1.553$, $d=0.189\pm0.003$, $e=-0.814\pm0.162$ and $f=2.849\pm0.02$. At small values of $z$, it is clear that the dependence is linear, 
\begin{equation}\label{lineargofz2}
g(z) = c + {\cal C} z. 
\end{equation}
Obviously, the intersect $c$ is much smaller than the slope ${\cal C}$. The sign of $g(z)$ can be flipped regarding to the sign of its independent variable $z$. Accordingly, we get  
\begin{equation}\label{lineargofz}
g(z)\approx {\cal C} z. 
\end{equation}

To prove this dependence, algebraically, we try to estimate $g(z)$ directly from the division of $F_0$ by $F_1$, which can be approximated by including their first terms only, i.e. 
\begin{equation} \label{eq.A1}
g(H)\approx \frac{3(\gamma-2)}{2[1+(1-r)\gamma]}\; H^{1-r},
\end{equation}
Then, we approximate $z(H)$ to the form,
\begin{equation} \label{eq.A2}
z(H)\approx-\frac{3[1+(1-r)\gamma]}{2(1-r)}\; H^{1-r}.
\end{equation}
Finally, we now able to derive an approximate estimation for $g(z)$. According to Eq.~(\ref{eq.A1}) and (\ref{eq.A2}), we get
\begin{equation}
g(z)\approx \frac{(1-r)(\gamma-2)}{[1+(1-r)\gamma]^2}\; z
\end{equation}
Amazingly, this expression looks the same as the one we obtained from the numerical approximation with 
\begin{equation}
 {\cal C} = \frac{(1-r)(\gamma-2)}{\left[1+(1-r)\gamma\right]^2}. 
\end{equation}

\section*{Appendix C: Solution of Abel equation $y\dot y -y = ax$} \label{App:B}

To solve Eq.~(\ref{abel1}) we divide the whole equation by $y^3$ and introduce
a new variable $v=1/y$. Then Eq.~(\ref{abel1}) reads
\begin{equation}
\frac{dv}{dx}+v^{2}+axv^{3}=0.
\end{equation}
We then introduce the function $v=w/x$.
\begin{equation}
x\frac{dw}{dx}=w-w^{2}-aw^{3},  \label{abel2}
\end{equation}
Previous differential equation can be solved by separation of variables
\begin{equation}
\int \frac{dw}{w-w^{2}-aw^{3}}=\ln C^{-1}x,
\end{equation}
where $C$ is an arbitrary constant of integration. To calculate the
integral, we write the function to be integrated as
\begin{equation}
\frac{1}{w-w^{2}-aw^{3}}=\frac{1}{w}-\frac{aw}{aw^{2}+w-1}-\frac{1}{%
aw^{2}+w-1}.
\end{equation}
Let us assume that $\Delta =1+4a>0$ (this implies that $a>0$).
\begin{equation}
\int \frac{dw}{w-w^{2}-aw^{3}}=-\frac{1}{2\sqrt{\Delta }}\ln \frac{2aw-\sqrt{%
\Delta }+1}{2aw+\sqrt{\Delta }+1}-\frac{1}{2}\ln \left( aw^{2}+w-1\right)
+\ln w.
\end{equation}

Therefore the general solution of Eq. (\ref{abel2}) can be written as
\begin{equation}
x=C\frac{w}{\sqrt{aw^{2}+w-1}}\left( \frac{2aw+\sqrt{\Delta }+1}{2aw-\sqrt{%
\Delta }+1}\right) ^{1/2\sqrt{\Delta }},
\end{equation}
leading to
\begin{equation}
y=\frac{1}{v}=\frac{x}{w}=C\frac{1}{\sqrt{aw^{2}+w-1}}\left( \frac{2aw+\sqrt{%
\Delta }+1}{2aw-\sqrt{\Delta }+1}\right) ^{1/2\sqrt{\Delta }}.
\end{equation}

\end{document}